\documentclass[conference,10pt]{IEEEtran}
\usepackage{graphicx, epstopdf}
\usepackage{amssymb}
\usepackage{amsmath}
\usepackage{amsthm}
\usepackage{mathrsfs}
\usepackage{cite}
\usepackage{color}
\usepackage[acronym]{glossaries}
\usepackage[T1]{fontenc}
\usepackage[latin9]{inputenc}
\usepackage{array}
\usepackage{textcomp}
\usepackage{multirow}
\usepackage{soul}
\usepackage{flushend}

\newtheorem{remark}{Remark}
\makeatletter
\newcommand*\titleheader[1]{\gdef\@titleheader{#1}}
\AtBeginDocument{%
  \let\st@red@title\@title
  \def\@title{%
    \bgroup\normalfont\small\centering\@titleheader\par\egroup
    \vskip0.5em\st@red@title}
}
\title{Green Cell-less Design for RF-Wireless Power Transfer Networks}
\author{\IEEEauthorblockN{Ha-Vu Tran\IEEEauthorrefmark{1} and Georges Kaddoum\IEEEauthorrefmark{1}}
\IEEEauthorblockA{\IEEEauthorrefmark{1}ETS Engineering School, University of Quebec,
Montreal, Canada\\ 
Email: ha-vu.tran.1@ens.etsmtl.ca, georges.kaddoum@etsmtl.ca
}}
\begin{document}
\makeatother
\titleheader{This is the authors'version of the paper that has been accepted for publication in IEEE Wireless Communications and Networking Conference, 15-18 April 2018, Barcelona, Spain}
%\rfoot{Personal use of this material is permitted. Permission from IEEE must be obtained for all other uses, in any current or future media, including reprinting/republishing this material for advertising or promotional purposes, creating new collective works, for resale or redistribution to servers or lists, or reuse of any copyrighted component of this work in other works.}

    \maketitle
\begin{abstract}
This paper studies a new concept so-called green cell-less radio frequency (RF) wireless power transfer (WPT) networks.  We consider a scenario in which multiple indoor access points (APs) equipped with outdoor energy harvesters are connected with a central control unit via backhaul links. Further, such APs exploit the harnessed green energy to recharge wirelessly indoor devices under the coordination of the control unit. Considering the network, we focus on AP selection and beamforming optimization to maximize the total energy harvesting (EH) rate. The resulting mathematical problem has the form of mixed-integer optimization that is intractable to solve. Thus, we propose an algorithm to tackle this difficulty.  Through numerical results, we show the advantages of the cell-less design over the conventional small-cell one to validate our ideas. In particular, the issue on safety requirements of human exposure to RF radiation is discussed. Finally, potential future research is provided.
\end{abstract}
\section{Introduction}
Nowadays, radio frequency (RF) wireless power transfer (WPT) has been considered as an engaging approach to prolong the lifetime of wireless devices \cite{Lu2015,Tabassum2015, HaVuTranPotentials}.
Its principle relies on the fact that radio signals belonging to a frequency range from 300 GHz to 3 kHz can be used to carry energy over the air. In RF-WPT networks, transmitters can proactively replenish energy-hungry devices by sending RF signals.
Over the last few years, the RF-WPT technique has been widely exploited in many interesting forms, such as wireless sensor networks, implanted body devices, and cellular networks \cite{Lu2015,Tabassum2015,SWIPTKaddoum}.

It is well-known that one of the main challenges in designing RF-WPT networks is dealing with negative impacts of propagation loss. 
Many efforts have aimed at improving energy transfer efficiency \cite{HJu2014, Ng2014, Chingoska2016, TRTran2016}. 
In this concern, there exist two critically important issues on the environmental consequences and the human exposure to electromagnetic radiation, discussed as follows.  
The first issue can be explained by the prompt increase of the electricity cost due to wireless network operation  \cite{Mahapatra2016}. 
Specifically, generating sufficient power to achieve the networks' requirements causes a significant amount of CO$_2$ footprint \cite{Fehske2011}. 
%Particularly, the overall footprint of information and communication technology services is predicted to triple between 2007 and 2020 \cite{Fehske2011}. 
As a solution, energy harvesting (EH) has been seen as a key technique towards a future green world \cite{Lu2015}.
In principle, EH techniques harness green energy from natural sources, e.g., solar and wind, and then contribute to reducing the overall footprint in order to protect surrounding environments. Nevertheless, it is inconvenient for indoor devices able to harvest sufficient energy from natural sources. Thus, this motivates us to investigate potential scenarios to bridge green resources to indoor energy-hungry devices.
%that base stations (BSs) can harvest-and-store green energy and then use such energy to recharge devices wirelessly.

Regarding the second issue of RF radiation, in practice, there exist tight restrictions applied for increasing the transmit power at base stations (BSs) since the intensity of microwave radiation can harm the human health. Moreover, due to using low-power BSs, the indoor networks, such as femtocell and picocell networks, can fail to adapt EH requirements while ensuring the restrictions.
%In this concern, the applications of distributed antennas to wireless-powered networks have showed a big potential \cite{KwanNg2015, Derrick2016a}.
The concept of {\it cell-less/cell-free} networks, recently promoted for the fifth generation (5G) networks \cite{TaoHan2017, HienQuocNgo2017}, can be applied to open some propitious solutions. 
In fact, this concept is an incarnation of cooperative multipoint joint processing and distributed antenna systems \cite{HienQuocNgo2017, KwanNg2015, Derrick2016a}. 
%It is worth noting that the applications of distributed antennas to wireless-powered networks have shown a big potential \cite{KwanNg2015, Derrick2016a}.
In cell-less networks, a device does not associate with any BSs before it starts to transmit information. In addition, a device can be served by more than one BS. As a result, energy-hungry devices can benefit from multiple sources, and then have more chance to harvest sufficient energy while the transmit power restriction is respected. 
%Many papers have studied network MIMO  and DAS , and indicated that network MIMO and DAS may offer higher rates than colocated MIMO. However, these works did not consider the case of very large numbers of service antennas. Related works which use a similar system model as in our paper are. In these works, DAS with the use of many antennas
\begin{figure*}[t]
\centering
{\includegraphics[width=0.5\textwidth]{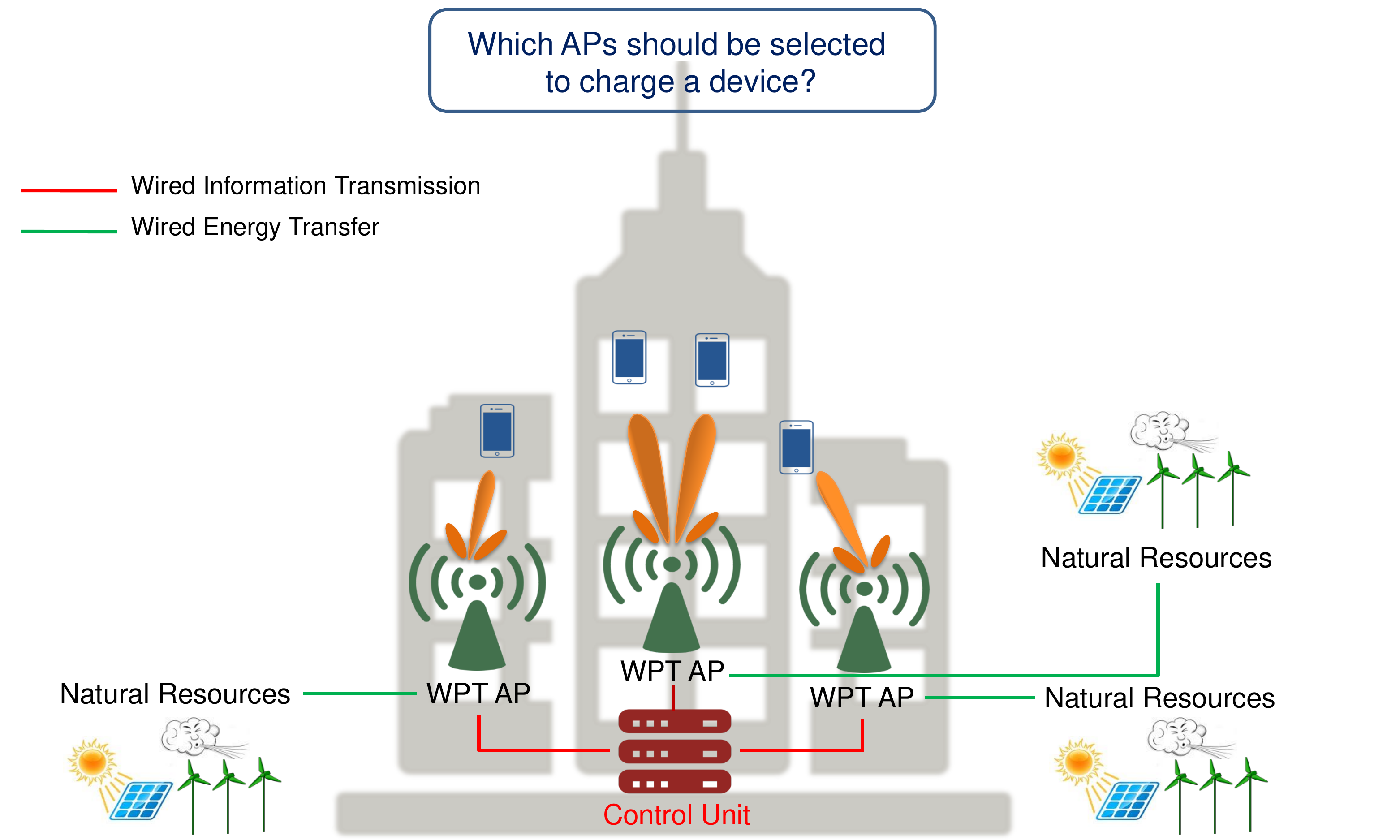}}
   \caption{
       An illustration of green cell-less RF-WPT networks.
    }
    \label{fig:SystemModel}
\end{figure*}

Based on the above discussion, we propose a novel application of the green cell-less design to RF WPT networks. 
In this model, access points (APs) are connected with outdoor energy harvesters, such as solar panels and wind turbines, to harvest-and-store green energy and then use such energy to recharge wirelessly multiple indoor devices using RF signals.
Additionally, all APs are connected with a control unit via backhaul links. This unit is responsible for managing communications in the networks, such as allocating resources.
Given this system, the appear of multiple APs can bring many benefits for EH. However, since the total transmit power is limited according to the transmit power restrictions and the collected natural energy at APs. Thus, this raises an interesting problem on AP selection and beamforming design to maximize EH performance. Therefore, in this work, we focus on the challenge that how the control unit optimally selects APs and allocates beamformers to serve each device such that the total EH rate is maximized. 
In this context, the resulting optimization problem has a mixed-integer formulation that is hard to solve \cite{AttahiruSAlfa2017}. To overcome this difficulty, we propose a low-complex method to tackle such a problem.

The main contributions of this work can be summarized by
\begin{itemize}
\item introducing the system model of green cell-less RF-WPT networks,
\item developing closed-form optimal solutions to solve the mixed-integer optimization problem regarding APs selection and beamforming optimization. 
\item discussing the obtained results based on safety requirements of human exposure to RF radiation.
\end{itemize}
%The remainder of this paper is organized as follows: In Section II, the system model is described. The issues on base station selection and beamforming design are discussed in Section III. In Section IV, numerical results are provided. Finally, concluding remarks are put forward in Section V.
%Specifically, wireless devices must satisfy the equivalent isotropically radiated power (EIRP) requirement ruled by Federal Communications Commission. For example, a maximum 36 dBm EIRP is limited on 2.4 GHz band. 
%In this concern, distributed antenna systems are an attractive technique to reduce electromagnetic pollutions in wireless powered networks [xx].

\section{System Model}
As shown in Fig. \ref{fig:SystemModel}, we consider an RF WPT network consisting of one control unit, $K$ WPT APs, and $N$ energy-hungry devices. Each device $n$ $(n = 1, 2, \ldots, N)$ is a single antenna device. Additionally, each WPT AP $k$ $(k = 1, 2, \ldots, K)$ is equipped with $M_k$ antennas and conventional energy harvesters, such as solar panels and wind turbines, to exploit natural energy. 
%Thus, it is assumed that each AP has a budget of transmit power, denoted by $\{ P^0_k \}$, depending on the available amount of green energy from outdoor energy harvesters. 
The APs are connected with the control unit via wired backhaul links. In particular, the control unit is responsible for computing and then forwarding resource allocation schemes to WPT APs.

Let ${\mathbf h}_n^k \in {\mathbb C}^{M_k \times 1}$ denotes the transmission channel between AP $k$ and device $n$. We assume each channel holds the reciprocity property. On this basis, the downlink channel state information (CSI) is estimated at the APs via a reverse time duplex division (TDD) scheme. Thus, through backhaul connections, all the CSI at each AP can be available at the control unit. 

Further, we define the variables $\{ \alpha_n^k \}$ ($\alpha_n^k \in \{0,1\}$) such that $\alpha_n^k = 1$ signifying AP $k$ is selected to serve device $n$. Hence, the received signal of device $n$ can be written as 
\begin{align}\label{eq:RSig}
y_n &= \sum\limits _{k=1}^{K} \sum\limits _{n'=1}^{N} (\alpha_{n'}^k{\mathbf w}_{n'}^{k})^T  {  {\mathbf h}_{n}^{k} } s_{n}^{k} + n_0,
\end{align}
where ${\mathbf w}_n^k \in {\mathbb C}^{M_k \times 1}$ is the beamforming vector used at AP $k$ for device $n$, $s_n^k$ is the energy signal with  unit power, and $ n_0$ is the additive white Gaussian noise (AWGN), i.e. $n_0 \sim \mathcal {CN} (0, \sigma^2_{0})$. 

Thus, the EH rate at device $n$ from $K$ APs can be computed by
\begin{align}
 \mathtt {EH}_n = \xi_n \sum\limits _{k=1}^{K} \sum\limits _{n'=1}^{N} \left| \alpha_{n'}^k({\mathbf w}_{n'}^{k})^T  {  {\mathbf h}_{n}^{k} } \right|^2,
\end{align}
where $\xi_n$ $(0 \le \xi_n \le 1)$ is the energy conversion efficiency at device $n$.

\section{Problem Formulation and Proposed Solution}
In this section, we show the problem formulation, and then discuss potential solutions.
\subsection{Problem Formulation}
The main concern of this work is to address the issues on APs selection and beamforming optimization. In this context, we aim at maximizing the total EH rate. Additionally, the transmit power should satisfy the constraints regarding the restriction level and the power budget at APs. Based on these requirements, conventionally, the corresponding optimization problem can be given by \footnote {It is worth noting that the extensions regarding simultaneous information and power transfer and dynamic EH from natural sources would be addressed in our future papers.}
\begin{subequations}\label{eq:MOProblem0}
\begin{align}
%\begin{aligned}
	\text{OP$_1$:} \quad &  \underset{{\bf w}^k_n, \alpha^k_n} \max \quad  \sum_{n=1}^{N} \xi_n \sum\limits _{k=1}^{K} \sum\limits _{n'=1}^{N} \left| \alpha_{n'}^k({\mathbf w}_{n'}^{k})^T  {  {\mathbf h}_{n}^{k} } \right|^2  \label{eq:MOProblema} 
\\
\text{s.t.:}  	\quad & \sum_{n=1}^{N} \left\| {{\bf w}^k_n} \right\|^2 \le \min \{P^1_k, P^0_k\}, \quad (\forall k,n) \label{eq:MOProblemb}
\\			
			\quad & \alpha_n^k \in \{0,1\}, \quad (\forall k,n), \label{eq:MOProblemc}
%\end{aligned}
\end{align}
\end{subequations}
where $\{ P_k^1 \}$ is the transmit power restriction applied at AP $k$, and $\{ P^0_k \}$ is a budget of transmit power at AP $k$ depending on the available amount of green energy from outdoor energy harvesters.
Accounting for problem OP$_1$, it is observed that it belongs to the class of mixed-integer optimization. Although such a problem can be tackled using the conventional branch-and-bound manner \cite{JClausen1999}, this might bring a heavy computational burden to the control unit.
To deal with this, our low-complex optimal solution is presented in the next subsection.
\subsection{Proposed Solution}
The main difficulty of solving OP$_1$ is dealing with the multiplication of two variables, i.e. $\alpha_{n}^k({\mathbf w}_{n}^{k})^T$, in which variables $\{ \alpha_n^k \}$ are integers. Luckily, it can be overcome based on a re-thinking of the problem's characteristics. Indeed, specifically, one can see that the objective function is non-decreasing. In addition, if we set variables $\{ \alpha_n^k \}$ holding "0" to "1", the optimal value of the problem does not decrease.
This can be explained by the natural characteristics of the maximization problems and the non-decreasing objective functions. 
Therefore, there exists an optimal solution of OP$_1$ for which $\{ \alpha_n^k \} = 1$ ($\forall n, k$). Based on this discussion, the solving algorithm of OP$_1$ can be divided into two following steps.
\subsubsection{Beamforming Design}
In the first step, by setting $\{ \alpha_n^k \} = 1$ ($\forall n, k$), problem OP$_1$ can be reformulated as 
\begin{subequations}\label{eq:MOProblem2}
\begin{align}
%\begin{aligned}
\text{OP$_2$:} \quad &  \underset{{\bf w}^k_n} \max \quad \sum_{n=1}^{N}  \sum\limits _{k=1}^{K} \sum\limits _{n'=1}^{N} \left| ({\mathbf w}_{n'}^{k})^T  {  {\mathbf h}_{n}^{k} } \right|^2  \label{eq:MOProblem2a} 
\\
\text{s.t.:}  	\quad & \sum_{n=1}^{N} \left\| {{\bf w}^k_n} \right\|^2 \le \min \{P^1_k, P^0_k\}. \label{eq:MOProblem2b}
%\end{aligned}
\end{align}
\end{subequations}
It is worth noting that constants $\{ \xi_n \}$ are omitted since they do not affect the optimal solutions. 

We denote $\{{\bf w}^{\star k}_n\}$ as the optimal solutions of OP$_2$.
Further, one can observe that problem OP$_2$ has the form of maximizing a convex function, unfavorable to solvers. 
%and then it is can be tackled by using solvers, such as CVX \cite{Gra2009}. 
To reduce the computational burden at the control unit, we endeavor to investigate the closed-form expression of $\{{\bf w}^{\star k}_n\}$.

Considering the objective function of OP$_2$, we can transform it into the following form
%\begin{footnotesize}
\begin{align}
\sum_{n=1}^{N} \sum\limits _{k=1}^{K} \sum\limits _{n'=1}^{N} \left| ({\mathbf w}_{n'}^{k})^T  {  {\mathbf h}_{n}^{k} } \right|^2 = \sum\limits _{k=1}^{K} \sum_{n=1}^{N} \sum\limits _{n'=1}^{N} \left| ({\mathbf w}_{n}^{k})^T  {  {\mathbf h}_{n'}^{k} } \right|^2.
\end{align}
%\end{footnotesize}

Further, since OP$_2$ is a maximization problem and the transmit power at each AP is limited, we can decompose OP$_2$ into $K$ sub-problems without the loss of optimality. Specifically, the sub-problem for AP $k$ can be given by
\begin{subequations}\label{eq:subMOProblem}
\begin{align}
%\begin{aligned}
	\text{SubOP$_k$:} \quad &  \underset{\{{\bf w}^k_n\}_n} \max \quad  \sum_{n=1}^{N} \sum\limits _{n'=1}^{N} \left| ({\mathbf w}_{n}^{k})^T  {  {\mathbf h}_{n'}^{k} } \right|^2  \label{eq:subMOProblem2a1} 
\\
\text{s.t.:}  	\quad & \sum_{n=1}^{N} \left\| {{\bf w}^k_n} \right\|^2 \le \min \{P^1_k, P^0_k\}. \label{eq:subMOProblem2b1}
%\end{aligned}
\end{align}
\end{subequations}

The objective function \eqref{eq:subMOProblem2a1} can be presented as
\begin{align}
\sum_{n=1}^{N} \sum\limits _{n'=1}^{N} \left| ({\mathbf w}_{n}^{k})^T  {  {\mathbf h}_{n'}^{k} } \right|^2 = \sum_{n=1}^{N}   ({\mathbf w}_{n}^{k})^H \left( \sum\limits _{n'=1}^{N} {  {\mathbf h}_{n'}^{k} } ({  {\mathbf h}_{n'}^{k} })^H \right){\mathbf w}_{n}^{k}.
\end{align}
Interestingly, one can see that all optimal beamformers $\{{\bf w}^{\star k}_n\}_n$ would have the same beam direction (i.e. $\frac{\{{\bf w}^{\star k}_n\}}{ \left\| {{\bf w}^{\star k}_n} \right\|^2} \equiv \frac{\{{\bf w}^{\star k}_{n'}\}}{ \left\| {{\bf w}^{\star k}_{n'}} \right\|^2}$ $\forall n,n'$) since they all aim at maximizing the multiplication between them and $\left( \sum\limits _{n'=1}^{N} {  {\mathbf h}_{n'}^{k} } ({  {\mathbf h}_{n'}^{k} })^H \right)$. This implies that the AP should select the device with the best channel condition to convey energy signals with the maximal transmit power. Particularly, the optimal beamformer should be designed such that the total energy harvested at the desired device and the others is maximized. 

Accordingly, the selected device, indexed by device ${\bar n}$, can be found by the following rule
\begin{align}
{\bar n} =\arg\max\limits _{n }\{\{ |{\mathbf h}_{n}^{k}| \}_{n=1}^{N}\}.
\end{align}

%\begin{align}
%{\mathbf w}_{}^{\star k} = \sqrt{\min \{P^1_k, P^0_k\}}  {\mathbf v}_{}^k,
%\end{align}
%where ${\mathbf v}_{}^k$ is the orthonormal eigenvector corresponding to the largest eigenvalue $\lambda_\text{largest} \left( \sum\limits _{n'=1}^{N} {  {\mathbf h}_{n'}^{k} } ({  {\mathbf h}_{n'}^{k} })^T \right)$.

%Next, one can evaluate that the value of objective function \eqref{eq:subMOProblem2a1} achieves maximum when total transmit power is assigned for the device with the largest channel gains. 

%After that, the AP forms only one beam with total transmit power steering energy signals towards device ${\bar n}$. In other word, the optimal beamformer of AP $k$ is the solution of the following problem
%\begin{subequations}
%\begin{align}
%  \underset{{\bf w}^k_{\bar n}} \max \quad  &  \left| ({\mathbf w}_{\bar n}^{k})^T  {  {\mathbf h}_{\bar n}^{k} } \right|^2  \label{eq:subMOProblem2a} 
%\\
%\text{s.t.:}  	\quad & \left\| {{\bf w}^k_{\bar n}} \right\|^2 \le \min \{P^1_k, P^0_k\}. \label{eq:subMOProblem2b}
%\end{align}
%\end{subequations}

Thus, each AP only needs to form one beam with maximal transmit power. The closed-form solution ${\mathbf w}_{ n}^{\star k}$ can be shown as follows 
\begin{align}\label{eq:subMOProblem2}
{\mathbf w}_{ n}^{\star k}  = \left\{ {\begin{array}{*{20}{l}}
{\mathbf w}_{\bar n}^{\star k} \quad &\text{if} \quad n = {\bar n},  \\
0  \quad &\text{if} \quad \text{otherwise},
\end{array}} \right.
\end{align}
in which
\begin{align}
{\mathbf w}_{\bar n}^{\star k} = \sqrt{\min \{P^1_k, P^0_k\}}  {\mathbf v}_{\bar n}^k,
\end{align}
where ${\mathbf v}_{\bar n}^k$ is the orthonormal eigenvector corresponding to the largest eigenvalue $\lambda_\text{largest} \left( \sum\limits _{n'=1}^{N} {  {\mathbf h}_{n'}^{k} } ({  {\mathbf h}_{n'}^{k} })^H \right)$.

\subsubsection{Access Point Selection}
In the second step, we can find the optimal values of $\{ \alpha_n^k \}$, denoted by $\{ \alpha_n^{\star k} \}$, through a simple closed-form expression as below
\begin{align}\label{eq:DP1}
 \alpha_n^{\star k}  = \left\{ {\begin{array}{*{20}{l}}
1 \quad &\text{if} \quad \left\|{\bf w}^{\star k}_n\right\|^2 > 0,  \\
0  \quad &\text{if} \quad \text{otherwise}.
\end{array}} \right.
\end{align}
%Nevertheless, this procedure can become more flexible for network operators by the relaxed one as follows
%\begin{align}\label{eq:DP2}
% \alpha_n^{\star k}  = \left\{ {\begin{array}{*{20}{l}}
%1 \quad &\text{if} \quad \left\|{\bf w}^{\star k}_n\right\|^2 > \epsilon,  \\
%0  \quad &\text{if} \quad \text{otherwise},
%\end{array}} \right.
%\end{align}
%where $\epsilon$ is a rounding threshold. 

\begin{figure}[!]
\centering
{\includegraphics[width=0.28\textwidth]{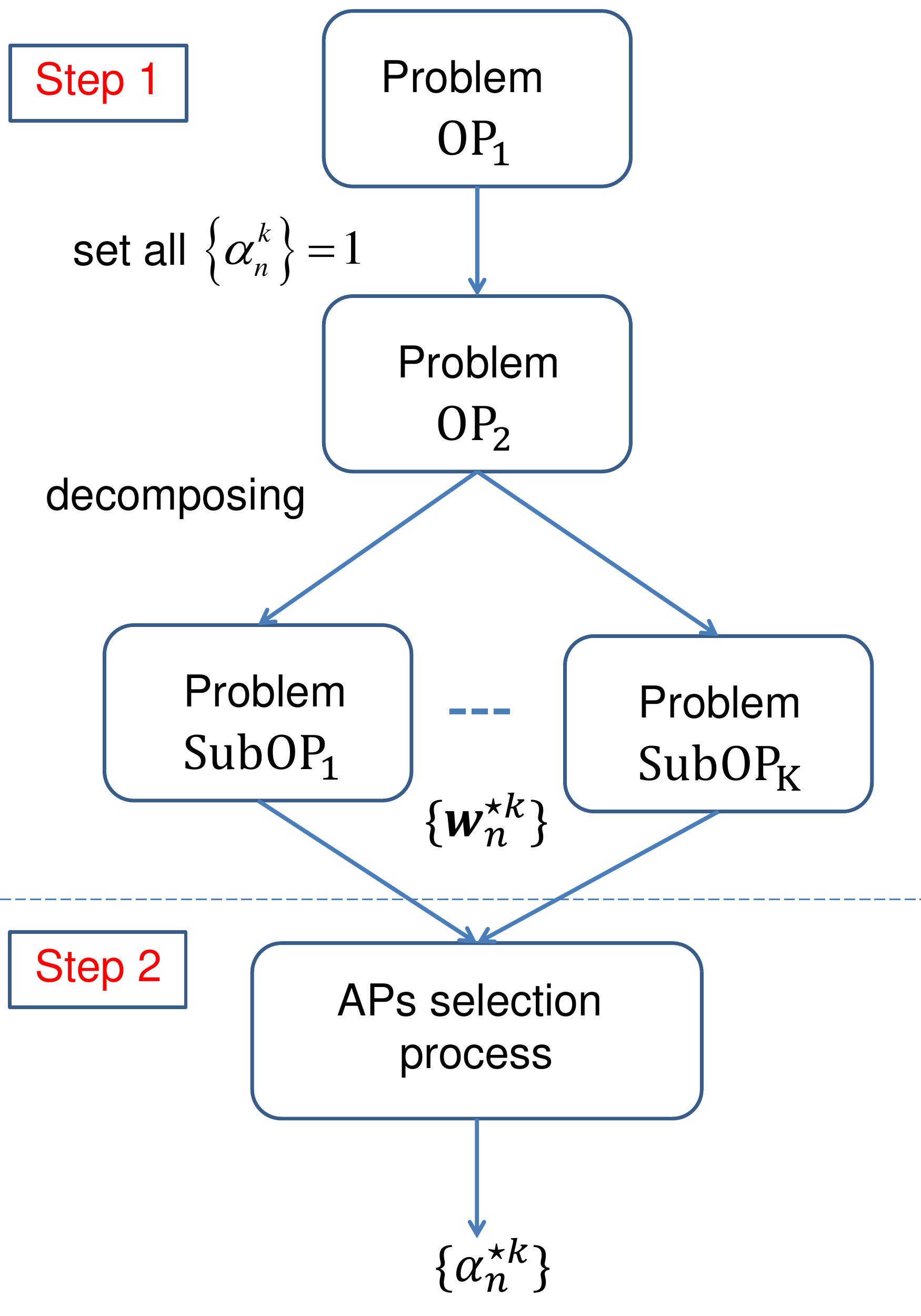}}
   \caption{
       The algorithm for AP selection and beamforming optimization.
    }
    \label{fig:Algo}
\end{figure}

For convenience, the flowchart of our algorithm is shown in Fig. \ref{fig:Algo}.
Using our algorithm, the main problem handling by the unit control can be split into $K$ subproblems. The unit control can solve these subproblems or distribute them to appropriate APs for further processes. Thus, the computation burden at the unit control can be relaxed. In addition, benefiting from the APs selection and beamforming, an AP can avoid putting the energy resource on inefficient transmissions to devices associated with limited channel gains.

\section{Health Safety Requirements and Charging Performance}
In this section, a discussion of health safety requirements and a calculation of charging performance are provided.

Regarding the safety levels for human body in a public environment, the permissible exposure level to RF electromagnetic fields are discussed in Remarks 1 and 2 as follows
\begin{remark}
Based on the IEEE Standard C95.1-2005, regarding the continuous and long-term exposure of all individuals, the basic restriction of the specific absorption rate (SAR) for frequencies between 100 kHz and 3 GHz is 0.08 W/kg averaged over any 6-minute-reference-period [pp. 20 \cite{IEEEstandard}]. This limit guarantees that there is no risk of adverse effects on the human health.
%It is worth noting that this limit is much lower than the SAR restriction applied for sensitive human parts, such as the eyes and brain, with localized RF exposure  [pp. 86--90 \cite{IEEEstandard}]. 
%regarding safety levels for human body in a public environment, the permissible exposure level to RF electromagnetic fields from 2 GHz to 100 GHz is 10 W/m$^2$ during 30 minutes. 
%Accordingly, since the total area of a smartphone (e.g. an iPhone) is roughly 0.008 m$^2$, the corresponding permissible exposure level applied on this area is 80 mW/0.008 m$^2$.
\end{remark}
Accordingly, for example of a 50 kg body mass, there is an SAR limit of 4 W averaged over a period of 6 minutes applied on the human body.

\begin{remark}
It is observed that the SAR limit depends on the body mass of the RF-absorbed people. The lower the mass is, the tighter the restriction is. For instance, the SAR restriction applied to children is tighter than the one applied to adults. Therefore, it is suggested future works should consider an upper limit for the EH rate on each typical user situation.
\end{remark}

Furthemore, the actual recharged energy obtained at a device can be computed by subtracting the consumed energy during recharging processs, i.e. in the supended state, from the total harvested energy.
Thus, the recharged battery percentage per hour can be obtained by the following calculation
\begin{align}\label{eq:charging} 
\mathtt {Per}_n = \frac{ \mathtt {ER}_n - \mathtt {ED}_n}{ \mathtt {V}_n} \times \frac{1}{\mathtt {BC}_n} \times 100,
\end{align}
where the rechared energy per hour $\mathtt {ER}_n$ (mW/h) can be computed as
\begin{align}
\mathtt {ER}_n&=\text {adapter efficiency} \times \mathtt {EH}_n \times 3600,
\end{align}
in which $\mathtt {Per}_n$, $\mathtt {ED}_n$ (mW/h), $\mathtt {BC}_n$ (mAH), and $\mathtt {V}_n$ (volts) represent the achieved battery percentage, energy discharged per hour, battery capacity, and voltage, respectively.

\section{Numerical Results}
\begin{table}
\begin{centering}
\caption{Important Parameters}
\par\end{centering}
\centering{}%
\begin{tabular}{|l|c|}
\hline
\textbf{Parameters} &
\textbf{System values}
\tabularnewline
\hline
Number of APs, $K$&
3\tabularnewline
\hline
Number of users, $N$ &
5\tabularnewline
\hline
Energy conversion efficiency, $\{ \xi \}$ &
50\% \tabularnewline
\hline
The budget of transmit power at each AP, $\{P^0_k\}$&
18 dBm \tabularnewline
\hline
The transmit power restriction at each AP, $\{P^1_k\}$&
20 dBm \tabularnewline
\hline
Rician factor&
10 dB \tabularnewline
\hline
Path loss exponent inside a building - line of sight&
1.7 \tabularnewline
\hline
\end{tabular}
\end{table}

\begin{figure}[t]
\centering
{\includegraphics[width=0.4\textwidth]{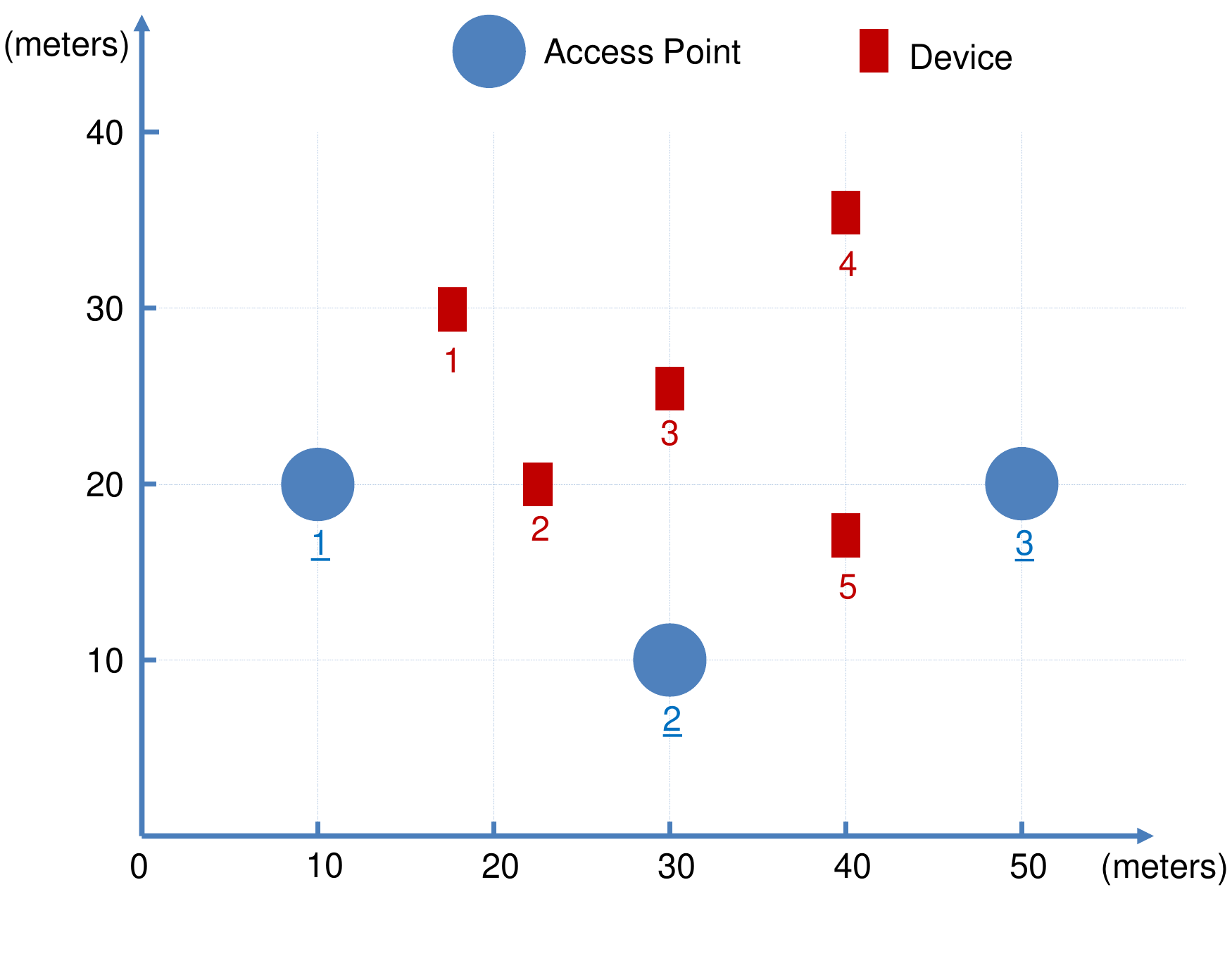}}
   \caption{
       A geographical illustration of the cell-less network model.
    }
    \label{fig:Sim1}
\end{figure}

\begin{figure}[!]
\centering
{\includegraphics[width=0.4\textwidth]{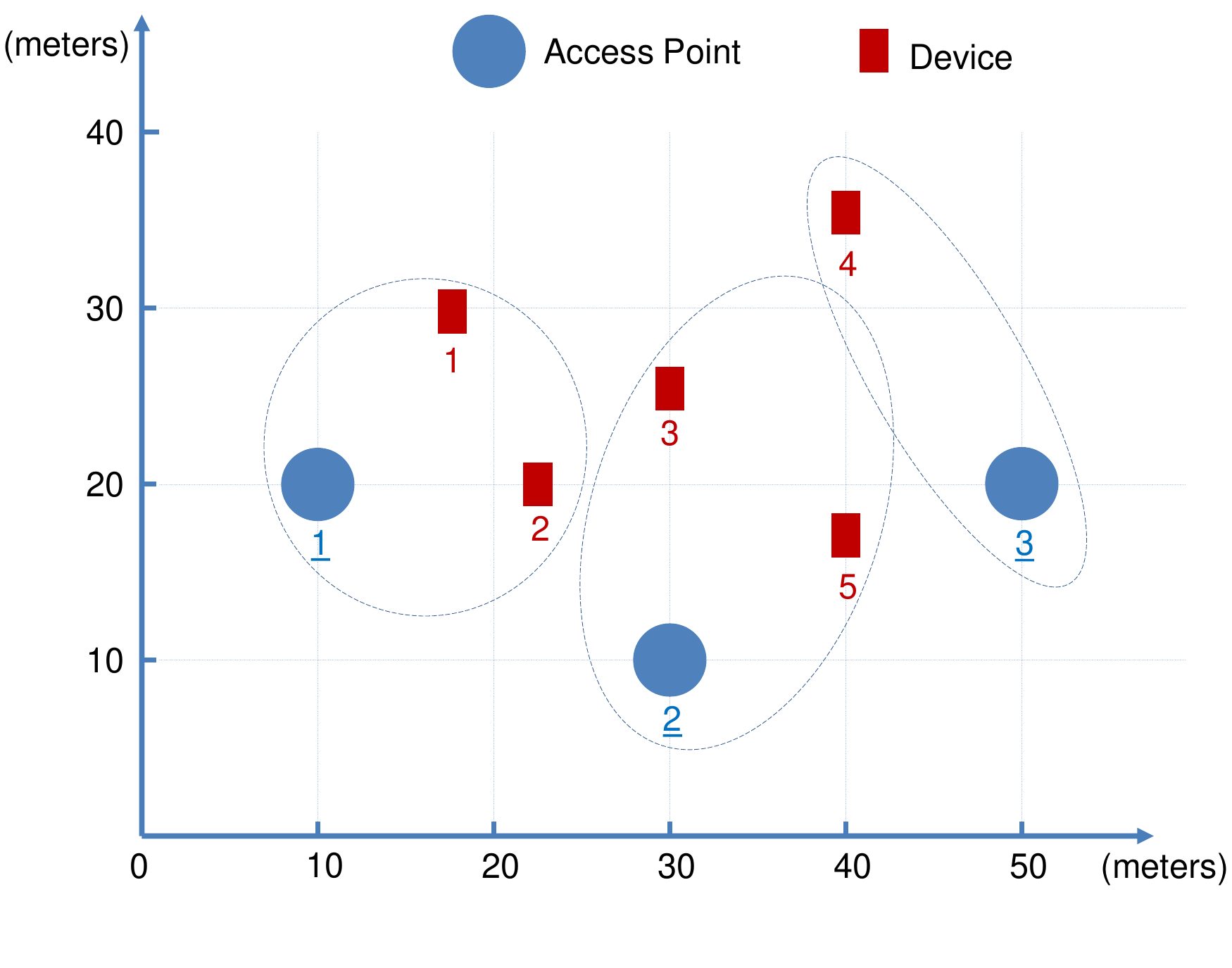}}
   \caption{
       A geographical illustration of a small-cell network model.
    }
    \label{fig:Sim2}
\end{figure}

This section is dedicated for showing numerical results in order to validate our ideas. 

In this simulation, we consider a WPT network consisting of 3 APs and 5 devices. Moreover, in order to compare the system performance between the cell-less network and the conventional small-cell one, we provide two geographical model versions, as shown in Fig. \ref{fig:Sim1} and Fig. \ref{fig:Sim2}. 
Furthermore, line of sight communication is considered in an office building, and the channels are assumed to follow Rician fading model. For more concerns, important system parameters are listed in Table 1. 

%\subsection{Cell-less WTP networks versus small-cell WPT networks} 
%\begin{figure}[t]
%\centering
%{\includegraphics[width=0.45\textwidth]{Images/Sim2.eps}}
%   \caption{
%    A comparison between solving approaches ($\{M_k\} = M = 3$).
%    }
%    \label{fig:Sim5}
%\end{figure}

\begin{figure}[t]
\centering
{\includegraphics[width=0.45\textwidth]{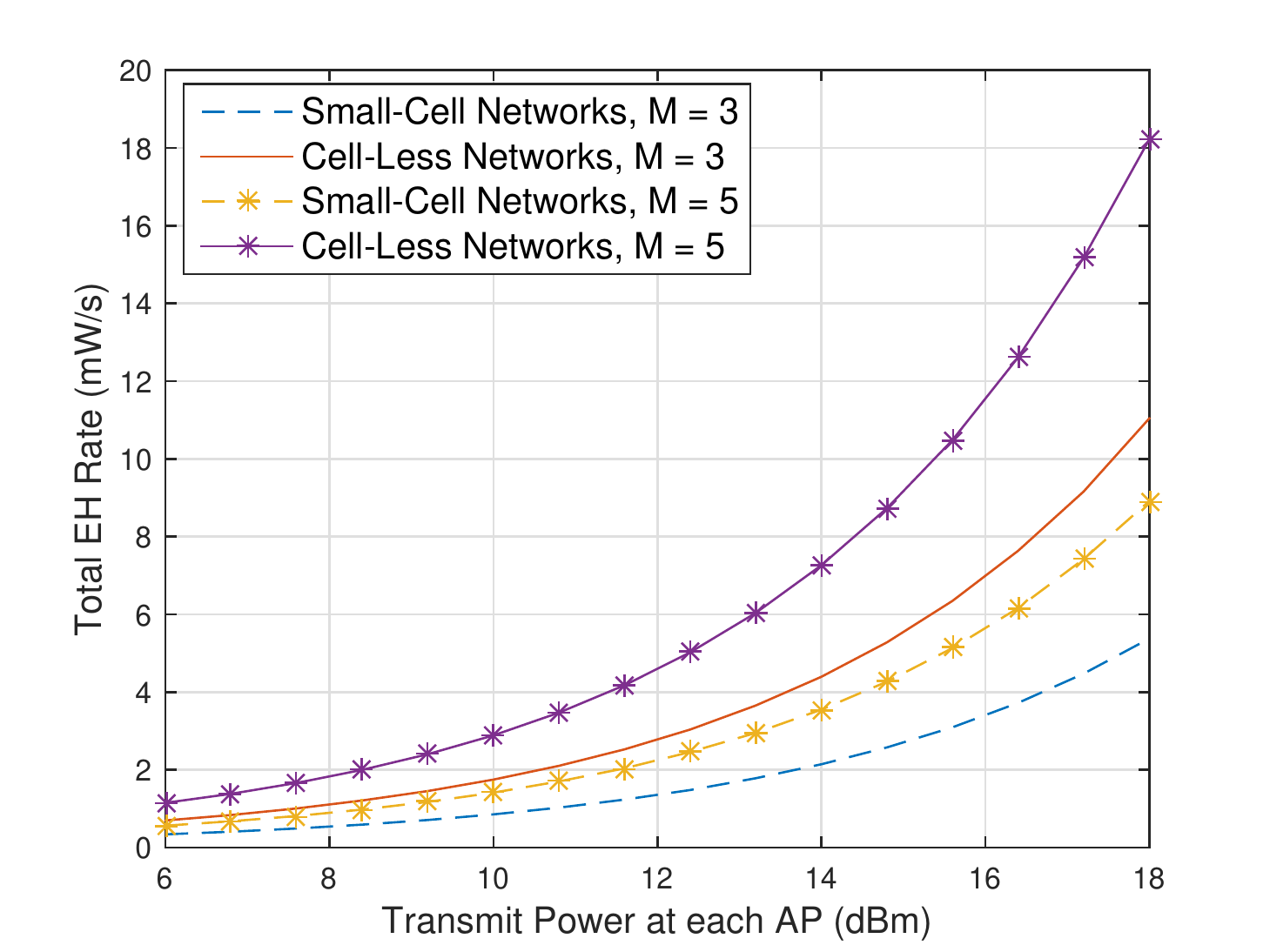}}
   \caption{
     The cell-less networks vs. the small-cell networks ($\{M_k\} = M$).
    }
    \label{fig:Sim3}
\end{figure}

%In this simulation, we start with demonstrating the advantages of APs selection and beamforming.
%In Fig. \ref{fig:Sim5}, a comparison between the proposed algorithm and semi-isotropic transmission (e.g. 180-degree radiation to the user side) is shown. It is visible that a big gain in terms of the total EH rate can be achieved when using our algorithm. It is due to that beamforming helps to focus the transmit energy signals while APs selection is to avoid transferring much power to the devices with weak channel gains.

In this simulation, we start with demonstrating the advantages of the cell-less design.
Fig. \ref{fig:Sim3} shows a performance comparison between the cell-less and the small-cell networks. In this context, the achievable EH rates of the both networks are used. 
The obtained numerical results reveal that the EH performance of the cell-less network is better than the one of the small-cell network. 
The outperformance of the cell-less network can be explained since the devices are not split into cells, more and nearer energy sources can be exploited.
For instance, when the value of the transmit power equals to 15 dBm, the performance gaps are approximate 5 mW/s and 3 mW/s in the cases of $M = 3$ and $M = 5$, respectively. Further, as observed in Fig. \ref{fig:Sim3}, the performance gap between the two approaches increases when the transmit power and the number of transmit antennas at each AP scale up.

\begin{figure}[t]
\centering
{\includegraphics[width=0.45\textwidth]{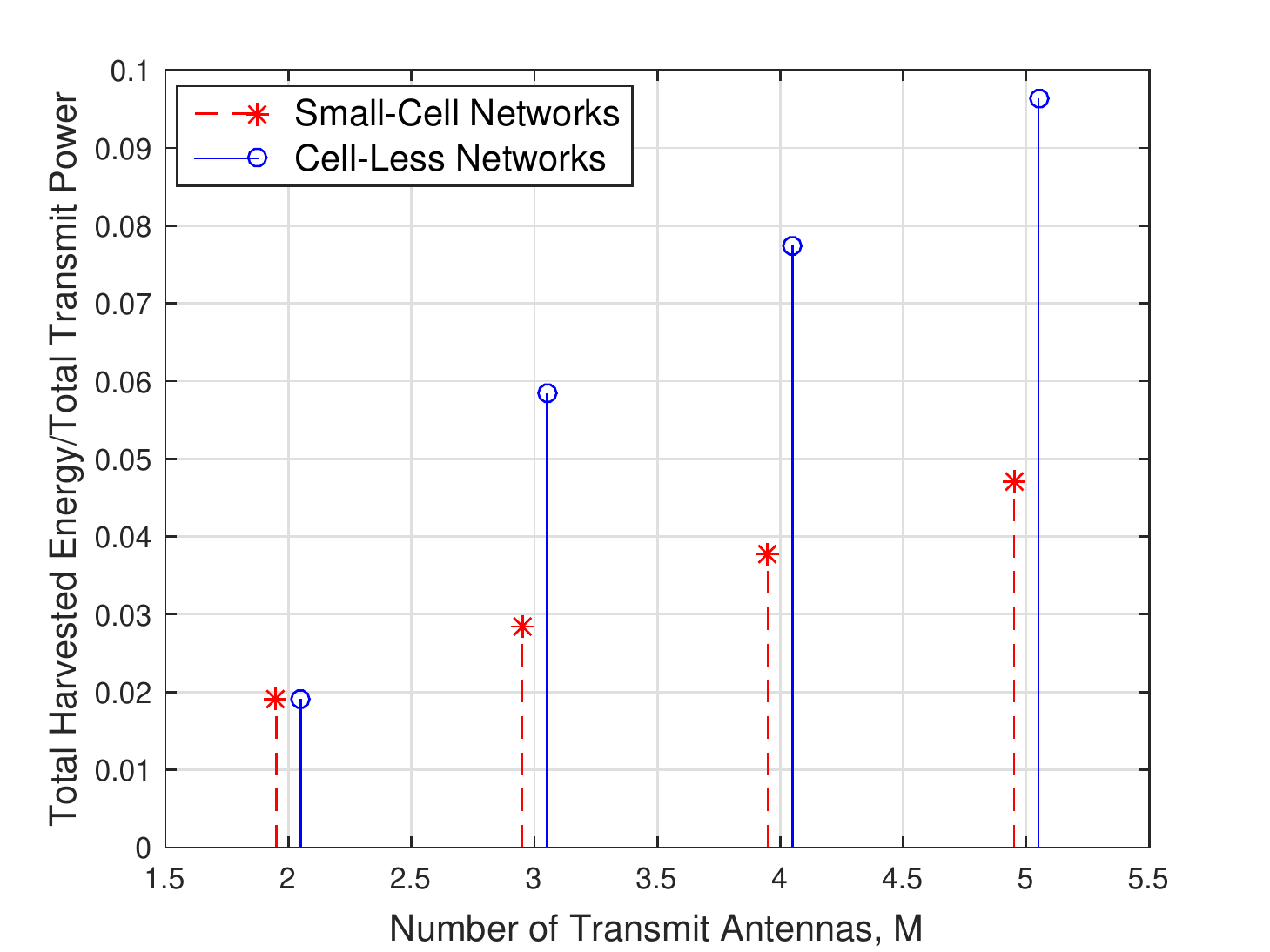}}
   \caption{
    An efficiency comparison between the networks.
    }
    \label{fig:Sim4}
\end{figure}

In Fig. \ref{fig:Sim4}, we evaluate the energy transfer efficiency of the two network concepts through the fraction between total harvested energy and total transmit power. Based on the comparison, the concept of the cell-less network might be more promising than the one of the small-cell network. It is observed that the energy transfer efficiencies and the performance gaps increase as the number of transmit antennas scale up. With APs selection and beamforming, the network can achieve up to 9\% transfer efficiency. Nowadays, TX91501 Powercaster$^\text{TM}$ transmitter can provide 1 $\mu$W for a receiver over a distance of 11 m by using 3 W equivalent isotropically radiated power (EIRP) \cite{TX91501}. This implies 3.33 $\times$ 10$^{-5}$ \% efficiency which is much lower than the obtained results. Therefore, the combination of beamforming, APs selection, and the cell-less structure can be evaluated as future trends for RF-WTP networks.

\begin{figure}[t]
\centering
{\includegraphics[width=0.45\textwidth]{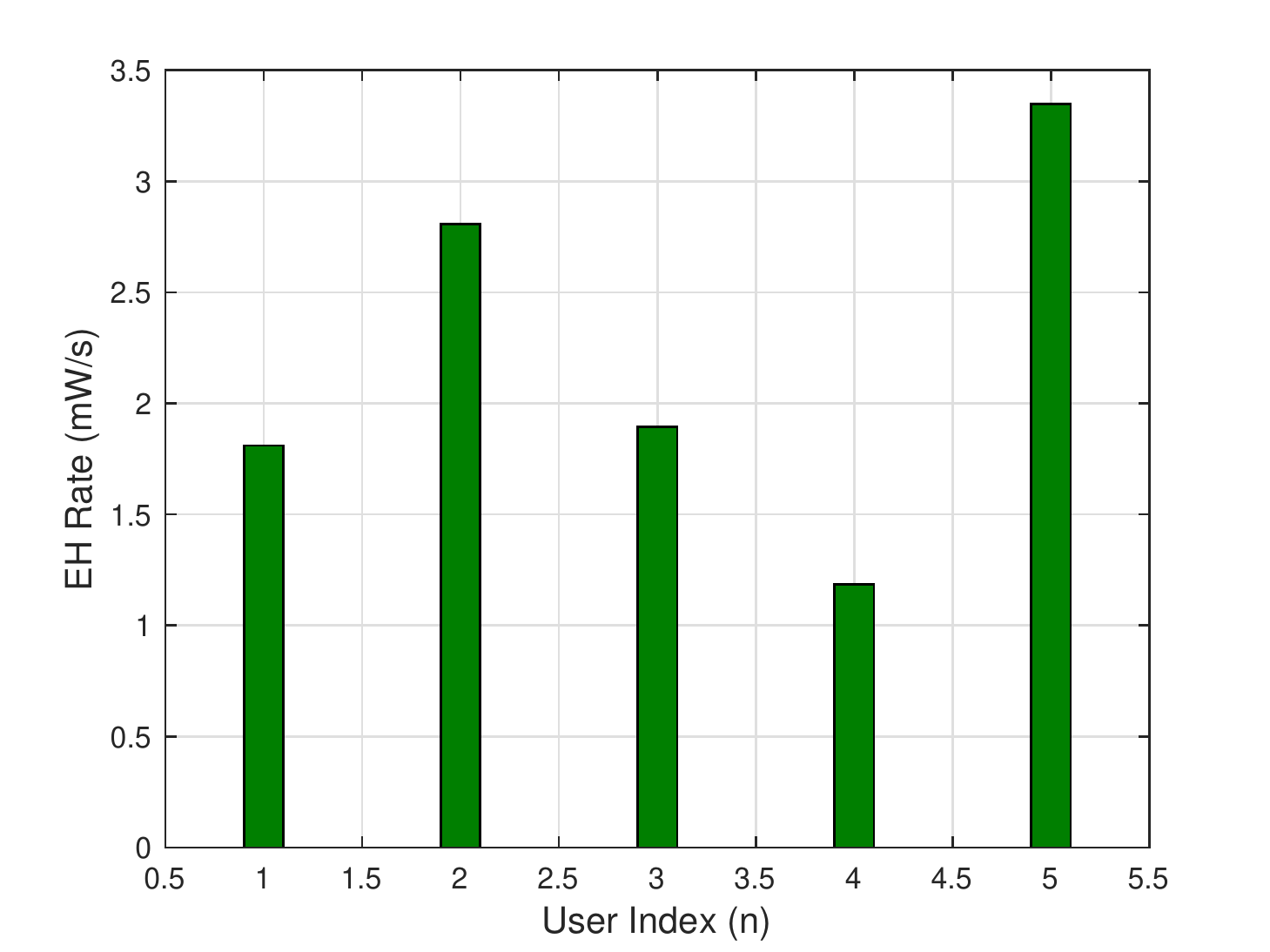}}
   \caption{
    Average EH rates achieved at each device.
    }
    \label{fig:Sim6}
\end{figure}

%\subsection{Health Safety Requirements and Charging Performance}

In Fig. \ref{fig:Sim6}, the average EH rates achieved at each device in the cell-less WPT networks are presented.
% Obviously, there is no position in the network suffering from the exposure that higher than the limit.
 Specifically, device 5 harvests the highest amount of energy since it is located in the most convenient position nearby APs. In this concern, equivalently, the total energy beamed at device 5 is approximately 6.694 mW/s (i.e. $\{ \xi_n \} = 50\%$) averaged over one second. We assume that the device location, i.e. a smartphone, is nearby its owner. Additionally, this person, whose mass is 50 kg, absorbs in the worst case the same amount of RF radiation as the device does. Then, the human exposure averaged over a period of 6 minutes can be computed by ${0.006694 \times 360} = 2.4098$ W which is lower than the limit according to Remark 1.

\begin{figure}[t]
\centering
{\includegraphics[width=0.45\textwidth]{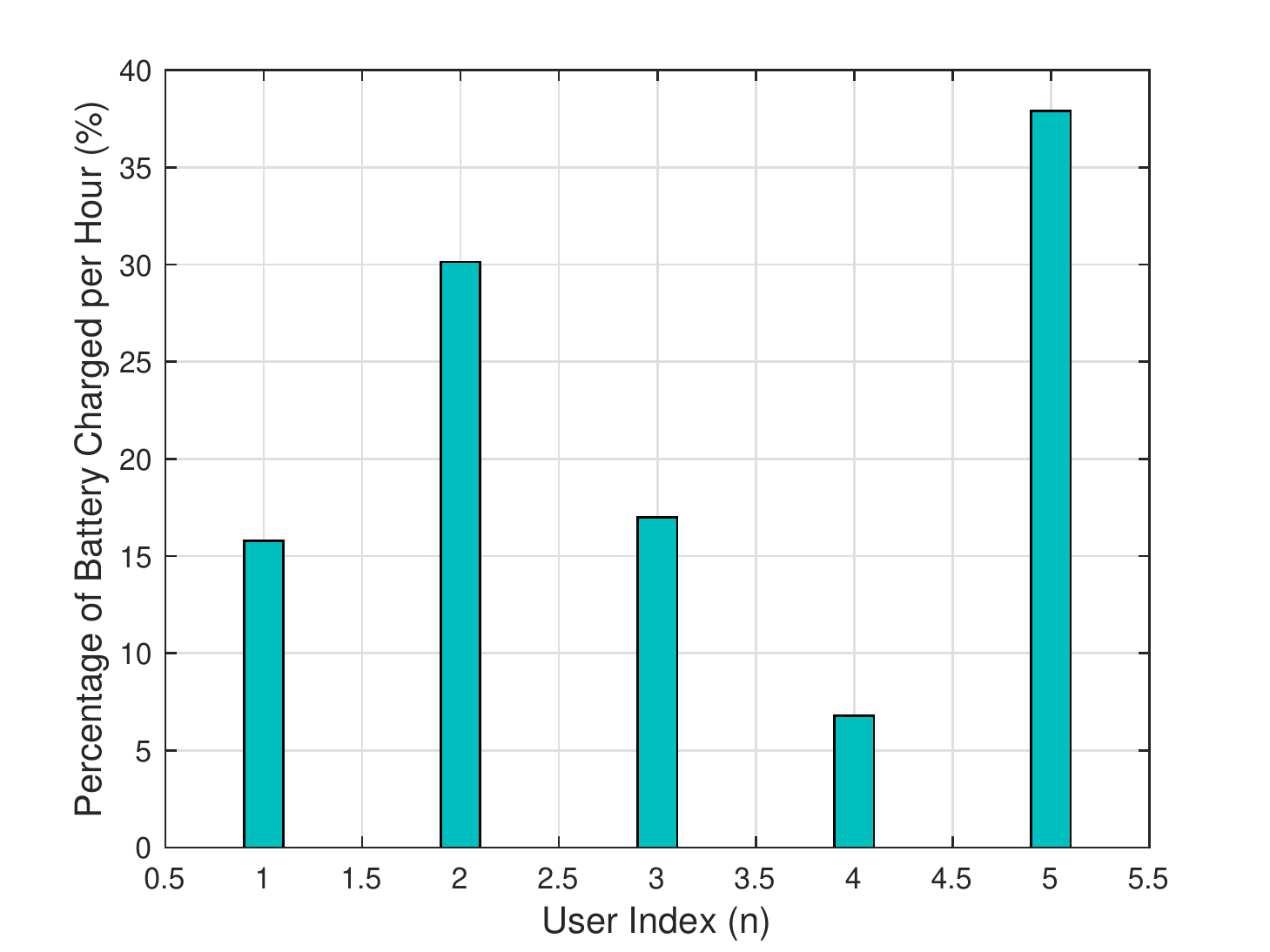}}
   \caption{
    Percentage of the battery capacity charged per hour at each device.
    }
    \label{fig:Sim7}
\end{figure}

Next, assuming that the battery capacity at each device is 4000 mAH, i.e. smartphones \cite{battery}, Fig. \ref{fig:Sim7} shows how many percent of the battery capacity at each device can be re-charged during 1 hour by using eq. \eqref{eq:charging}.
It is important to note that the voltage is 5 volts \cite{P2110B}, the adapter efficiency is 80\% \cite{Siddabattula}, and $\mathtt {ED}$ is 2058 mW \cite{Carroll2010}. 
In the best case, device 5 can be recharged up to 38\% of the battery capacity. In fact, this promising results reveal a bright future for the cell-less RF-WTP networks.
%However, in practice, the actual amount of recharged capacity mainly depends on the technologies of batterry storage (i.e. Lithium Ion (Li-ion), Nickel Cadmium (NiCd), Nickel Metal) which is out of scope of this paper. Thus, one can regard the given results as an upper-bound of the achievable performance at the users.

\section{Potential Future Research}
In fact, deploying the cell-less RF-WPT networks might imply many challenges. In the following, we discuss three interesting issues for future research.

First, in the cell-less RF-WTP networks where the channels are line of sight, and the distances between users are short, designing one dedicated beamfomer for each user can lead to inefficiently computing efforts. Thus, the scenarios in which multiple users are served by a wider beam can be considered as a key solution.
In this context, this yields the key challenge in handling an interesting trade-off between reducing the number of beamformers and increasing the beam widths.
Especially, in the case that only one beam at each AP is needed, it is suggested that analogue beamforming is one of the most cost-effective approach since this technique is simple and requires a minimal amount of hardware.

Second, jointly handling information transmission and power transfer in the cell-free networks is crucial. The main challenge behind this task is how to manage interferences since the networks are no longer {\it celled}. Further, designing resource allocation schemes implies the computational burden centralized at the control unit, which is a prominent issue. Therefore, this requires investigating approaches to share the burden between the unit and APs, as well as reduce the amount of signaling information sent via backhaul connections. 

Third, adaptive and cooperative scenarios for the cell-less RF-WTP networks need to be addressed.
For instance, there is a new user entering the network. It requests the network for battery recharging. So, how the network can re-act to this request while guaranteering a minimal influence to the other users?, which AP should be selected?, and are any additional beamformers needed? or do the APs only need to adjust existing beamformers (i.e. beam directions and beam widths) to re-charge the user? 

\section{Conclusion}
In this paper, we introduce a novel system model so-called the green cell-less WPT network. Particularly, we propose the resource allocation scheme in terms of APs selection and beamforming optimization to maximize the total EH rate of the network. The formulated problem belongs to the class of the mixed-integer optimization which is not easy to solve. By rethinking the problem's characteristics, this problem has been solved through a simple proposed algorithm and the closed-form optimal solutions. In particular, our algorithm can help the control unit to relax the computation burden. The numerical results indicate that the cell-less networks can provide a significant improved performance of EH compared with the small-cell ones while ensuring the safety requirements of the human health. Finally, the discussion of the obtained results following safety requirements has been provided to figure out potential future research.
\bibliographystyle{IEEEtran}
\bibliography{IEEEabrv,REF}
\end{document}